\documentclass[12pt]{iopart} 
\usepackage{graphicx}

\begin{document}

\title{Steady state of tapped granular polygons.}

\author{Carlos M Carlevaro$^{1,2}$ and Luis A Pugnaloni$^1$}
\address{$^1$ Instituto de F\'{\i}sica de L\'{\i}quidos y Sistemas Biol\'{o}gicos (CONICET La Plata, UNLP), Casilla de Correo 565, 1900, La Plata, Argentina.\\
$^2$ Universidad Tecnol\'ogica Nacional - FRBA, UDB F\'{\i}sica, Mozart 2300, C1407IVT Buenos Aires, Argentina.}
\ead{manuel@iflysib.unlp.edu.ar (C M Carlevaro)}

\begin{abstract}
The steady state packing fraction of a tapped granular bed is studied for different grain shapes via a discrete element method. Grains are monosized regular polygons, from triangles to icosagons. Comparisons with disk packings show that the steady state packing fraction as a function of the tapping intensity presents the same general trends in polygon packings. However, better packing fractions are obtained, as expected, for shapes that can tessellate the plane (triangles, squares and hexagons). In addition, we find a sharp transition for packings of polygons with more than 13 vertices signaled by a discontinuity in the packing fraction at a particular tapping intensity. Density fluctuations for most shapes are consistent with recent experimental findings in disk packing; however, a peculiar behavior is found for triangles and squares.
\end{abstract}

\maketitle

\section{Introduction}

Granular materials settle under gravity and come to mechanical equilibrium unless an external excitation is provided. The properties of such static packings are difficult to predict, since the history of preparation of the sample is important. However, there exist different protocols to prepare a granular bed in a well defined macroscopic state. In such state, the packing fraction (and other macroscopic observables such as the pressure on the container) are reproducible if the given protocol is followed. A canonical example of this is the steady states obtained by tapping the sample with a given intensity \cite{nowak}. After a suitable annealing, tapping at a constant intensity produces mechanically stable configurations (inherent states, or microstates) whose ensemble has well defined mean values of all macroscopic observables.

In recent years, the dependency of the steady state packing fraction, $\phi$, on the tapping intensity, $\Gamma$, has been shown to be nonmonotonic; presenting a minimum at relatively high values of $\Gamma$ for disks and spheres \cite{pugnaloni1,pugnaloni2}, and a maximum at very low $\Gamma$ for spheres \cite{rosato}. In general, the symbol $\Gamma$ is used for the reduced peak acceleration given to the system during a tap. However, we will use $\Gamma$ in what follows to refer to any suitable parameter that characterizes the tapping intensity.

On the one hand, there exist some studies on the response to tapping of non-spherical particles \cite{rankenburg,villarruel,lumay,ramaoli}, however these do not consider polygonal particles. On the other hand, the are some investigations on polygon packings \cite{cruz,limon,ammi}. These latter studies, however, do not focus on the steady state obtained after a repeated pulse excitation. Inspired by previous works on pentagon packings \cite{vidales1,vidales2}, we investigate the $\phi$--$\Gamma$ tapping curve in the steady state for monosized regular polygons with different number $N$ of vertices; from triangles ($N=3$) to icosagons ($N=20$). As the number of vertices grows, we expect polygon packings to approach the properties of disk packings. Since depending on the number of vertices these particles may or may not tessellate the plane, we also expect strong deviations from the general trends for some grain shapes. 

In this paper, we compare the general features found in the $\phi$--$\Gamma$ curve of disk packings with those of regular polygons. Although some general trends are conserved, new phenomenology emerges.

In Section \ref{simulation} we present the simulation technique and the model particles. In Section \ref{tesselation} we analyze the behavior of polygons with fewer than ten vertices. In Section \ref{transition} we present results for polygons of up to twenty vertices. Section \ref{fluctuations} is devoted to the study of the density fluctuations. Finally, we draw the conclusions in Section \ref{conclusions} and point out some interesting areas of research suggested by the new results.    

\section{Simulation}
\label{simulation}
We perform molecular dynamic type simulations by solving the Newton--Euler equations of motion for rigid bodies confined on a vertical plane. Gravity acts on the negative vertical direction. The bodies (particles) are placed in a rectangular box which is confined to move in the vertical direction. This box is high enough to avoid particles to contact the ceiling during the simulations. We prepare nineteen samples that consist of 500 monosized regular polygons of a single type (from triangles to icosagons) or monosized disks. Particles, initially placed at random without overlaps in the box, are let to settle until they come to rest in order to prepare the initial packing. Then, the same tapping protocol is applied to each sample. 

We set the particle--particle interactions to yield a normal restitution coefficient $\epsilon=0.058$ and a static and dynamic friction coefficient $\mu_s = \mu_d = 0.5$. The confining box is $24.8 r$ wide and $2000 r$ tall (with $r$ the radius of the particles). The particle--box friction coefficient is $\mu_s = \mu_d = 0.07$ and the restitution coefficient is as in the particle--particle interaction. All polygons have the same radius and material density. Therefore, the actual weight of a particle depends on the number of vertices. We use as unit mass, $m$, the mass of a disk; as unit length $r$; and the unit time is $(r/g)^{1/2}$, with $g$ the acceleration of gravity.  

Tapping is simulated by giving the box an impulse. In practice, we set the initial velocity $v_0$ of the box (originally at rest after deposition) to a given positive value and restart the dynamics. In doing so, the box and its filling move upward and fall back on top of a zero restitution base. While the box dissipate all its kinetic energy on contacting the base, particles inside the box bounce against the box walls and floor until they fully settle. After all particles come to rest a new tap is applied. The intensity of the taps is measured by the initial velocity imposed to the confining box at each tap (i.e. $\Gamma=v_0$). A similar parameter (the lift-off velocity) has been recently proposed as a suitable measure of the tap intensity \cite{dijksman}.

The tapping protocol consist in a series of $50000$ taps. Every $250$ taps we change the value of $\Gamma$ by a small amount $\Delta \Gamma$. We initially decrease $\Gamma$ from $\approx 15.0 (rg)^{1/2}$ down to a very low value and then increase it back to its initial high value. At each value of $\Gamma$ the last $150$ taps are used to average the packing fraction in order to plot the $\phi$--$\Gamma$ curve.

The simulations were implemented by means of the Box2D library \cite{box2d}. Box2D uses a constraint solver to handle hard bodies. At each time step of the dynamics a series of iterations (typically 20) are used to resolve penetrations between bodies through a Lagrange multiplier scheme \cite{catto}. After resolving penetrations, the inelastic collision at each contact (a contact is defined by a manifold in the case of polygons) is solved and new linear and angular velocities are assigned. The equations of motion are integrated through a symplectic Euler algorithm. The time step $\delta t$ used to integrate the equations of motion is $0.025 \sqrt{d/g}$. Solid friction is also handled by means of a Lagrange multiplier scheme that implements the Coulomb criterion. This library achieves a high performance when handling complex bodies such as polygons. 

\section{Results}

\subsection{From triangles to nonagons}
\label{tesselation}

\begin{figure}
 \centering
 \includegraphics[width=0.8\columnwidth]{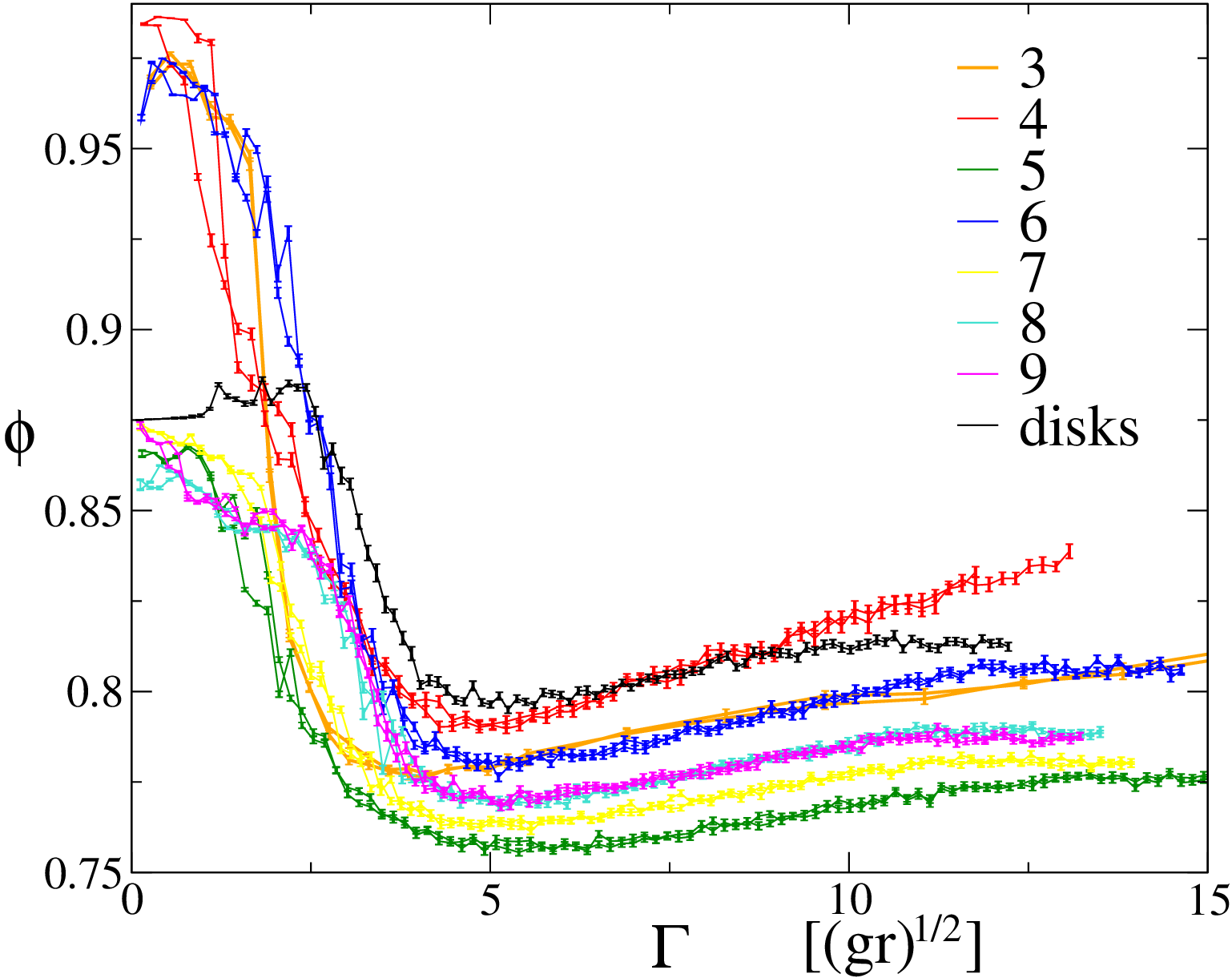}
 \caption{Mean packing fraction $\phi$ as a function of tapping intensity $\Gamma$ for triangles (violet), squares (red), pentagons (green), hexagons (blue), heptagons (yellow), octagons (cyan), nonagons (magenta), and disks (black). Except for disks, all curves correspond to a progressive decrease of $\Gamma$ followed by an increase back to high values. For disks only the decreasing part has been carried out. Error bars correspond to the estimated error of the mean.}
 \label{fig1}
\end{figure}

The steady state packing fraction as a function of the tapping intensity for triangles, squares, pentagons, hexagons, heptagons, octagons and nonagons is presented in Fig.~\ref{fig1} alongside with the results for disks. Packing fraction is estimated from the number density measured in a rectangular slab of half the packing hight at the middle of the sample. The fact that the same $\phi$-$\Gamma$ curve is obtained for decreasing and increasing $\Gamma$ indicates that these states are reversible and that $\phi$ is uniquely defined for each $\Gamma$. From this results we can see that polygon packings present similar features to those observed in disk packings. At low tapping intensities, a decrease of $\phi$ is observed for increasing $\Gamma$ down to a minimum packing fraction $\phi_{min}$. A further increase of $\Gamma$ induces an increase of $\phi$ until a plateau is reached at a packing fraction somewhat lower than the maximum obtained for the lowest values of $\Gamma$. Disks also show a not very pronounced maximum at low $\Gamma$ which is not observed in polygon packings. This maximum has been recently observed in sphere packings \cite{rosato}. Another overall trend is that the range of packing fractions attained by disk packings is narrower than for polygons.

Beyond these general features, there are some peculiarities associated to the ability of a given polygonal shape to tessellate the plane. As is to be expected, triangles, squares and hexagons can reach packing fractions of nearly $1$ at the lower tapping intensities. All other shapes reach packing fractions similar to disk packings at low $\Gamma$. It is important to notice at this point that our results differ from those obtained by Vidales \emph{et al.} \cite{vidales1,vidales2} in the case of pentagons in the framework of a pseudo-dynamic algorithm. In Refs. \cite{vidales1,vidales2} the $\phi$--$\Gamma$ curve does not present any minimum of the packing fraction.

\begin{figure}
 \centering
 \includegraphics[width=0.8\columnwidth]{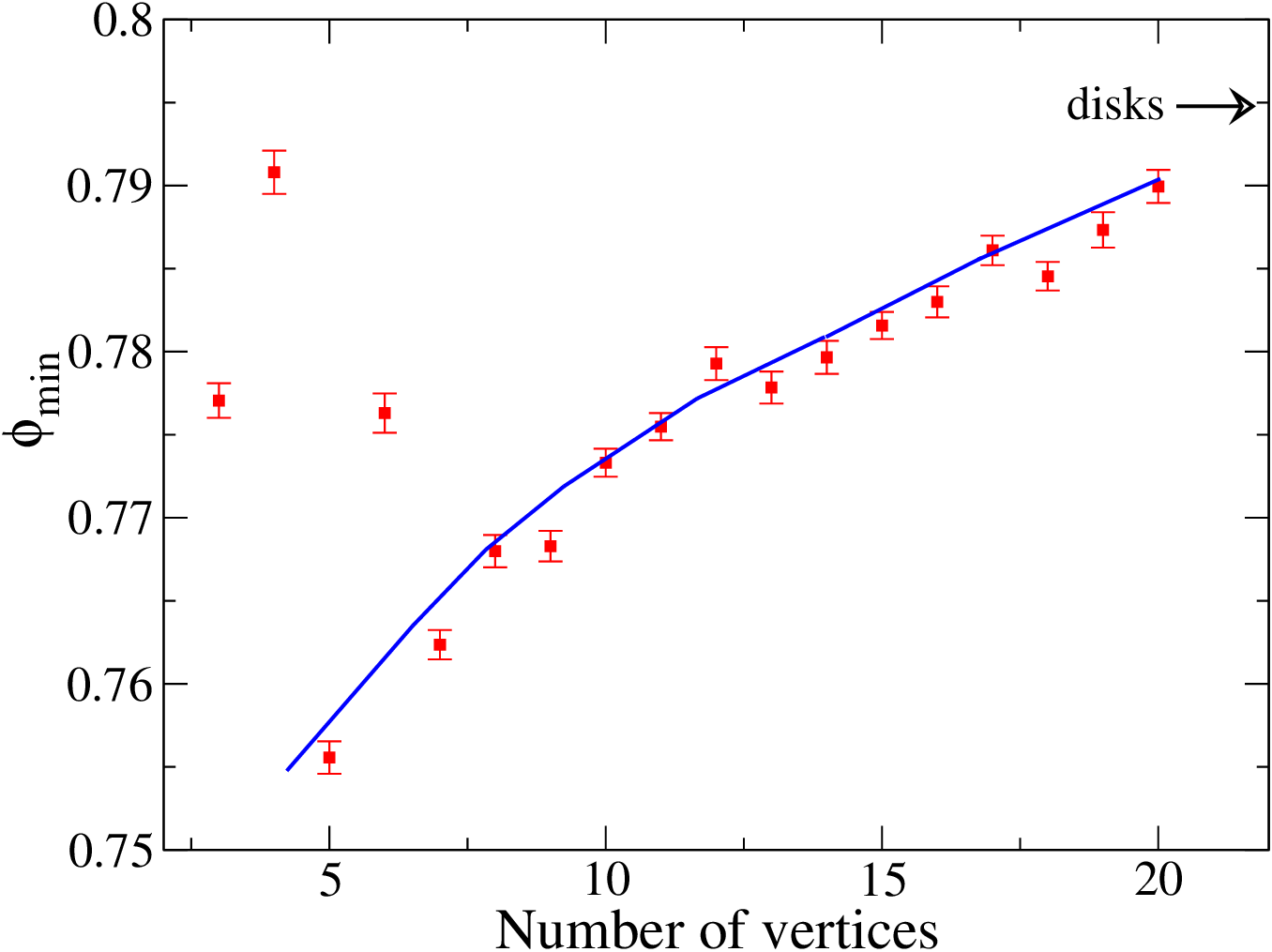}
 \caption{Minimum packing fraction $\phi_{min}$ as a function of the number of vertices. The blue line is drawn only to guide the eye.}
 \label{fig2}
\end{figure}

In Fig.~\ref{fig2}, we plot the minimum steady state density, $\phi_{min}$, as a function of the number of vertices of the polygon. As the number of vertices is increased, a consistent increase of $\phi_{min}$ is found for all polygons with the exception of triangles, squares and hexagons. As we mentioned, these three polygons can tessellate the plane. Correspondingly, triangles, squares and hexagons present higher densities than expected by the trend showed by all other polygons. We have seen that the position, $\Gamma_{min}$, of the minimum is independent of the number of vertices. The existence of $\phi_{min}$ has been associated to a competition between arch formation and arch breaking \cite{pugnaloni1}. The position $\Gamma_{min}$ of such minimum signals the crossover between a regime where arches cannot form due to the particles settling one by one (in a sequential manner) at very high $\Gamma$, and a regime where arches do form but are ``melted down'' in successive taps creating a dynamic equilibrium. The fact that $\Gamma_{min}$ is the same for all shapes is a clear indication that arching is not favored (nor prevented) by any particular shape at these intermediate values of $\Gamma$.  

\subsection{An unforeseen sharp transition for triskaidecagons and beyond}
\label{transition}

\begin{figure}
 \centering
 \includegraphics[width=0.9\columnwidth]{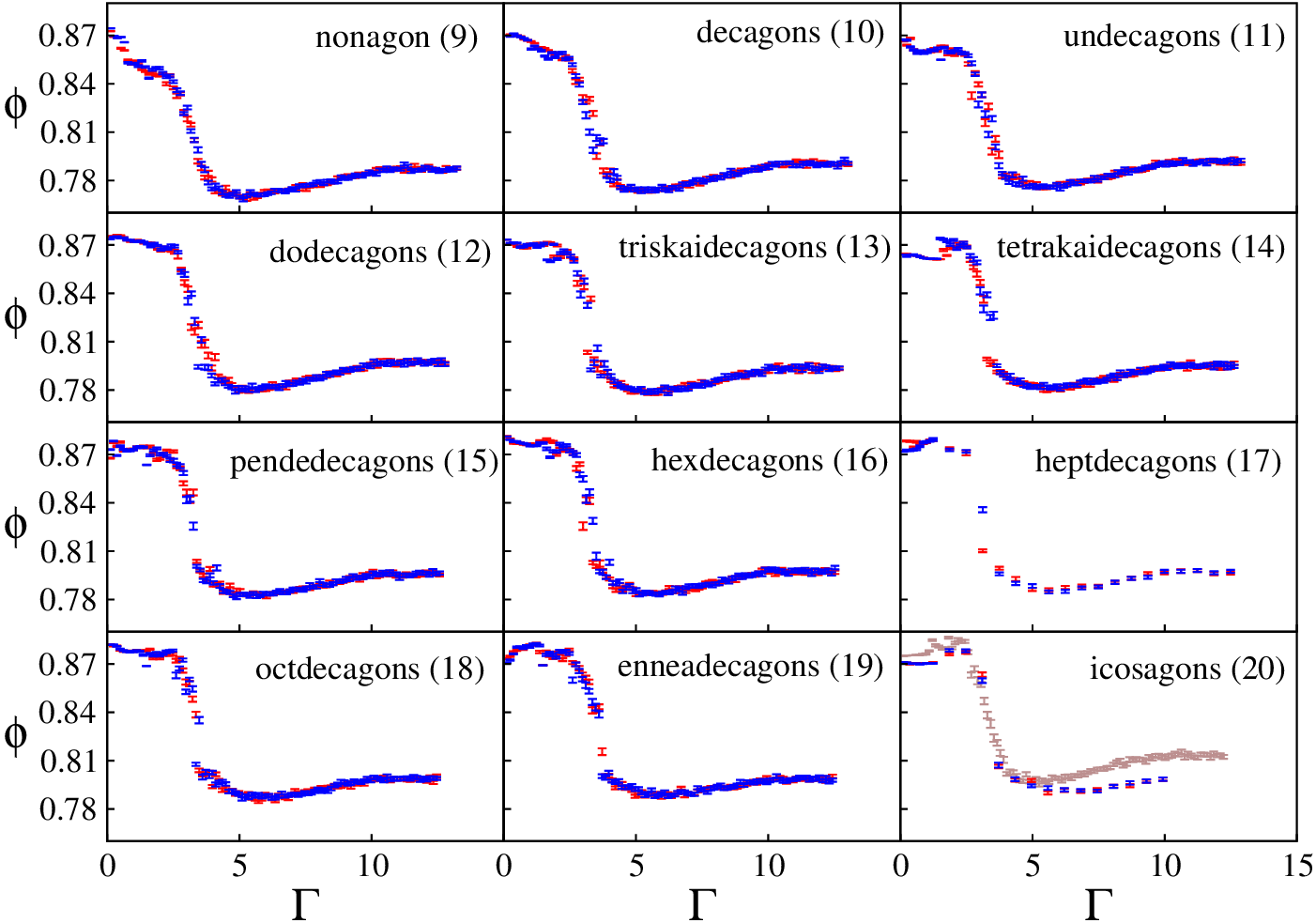}
 \caption{Mean packing fraction $\phi$ as a function of tapping intensity $\Gamma$ for nonagons, decagons, ... and icosagons. The brown data in the lower right panel correspond to disks. A progressive decrease (red data) of $\Gamma$ is followed by an increase (blue data) back to the high initial values. Error bars as in Fig.~\ref{fig1}.}
 \label{fig3}
\end{figure}

\begin{figure}
 \centering
 \includegraphics[width=0.9\columnwidth]{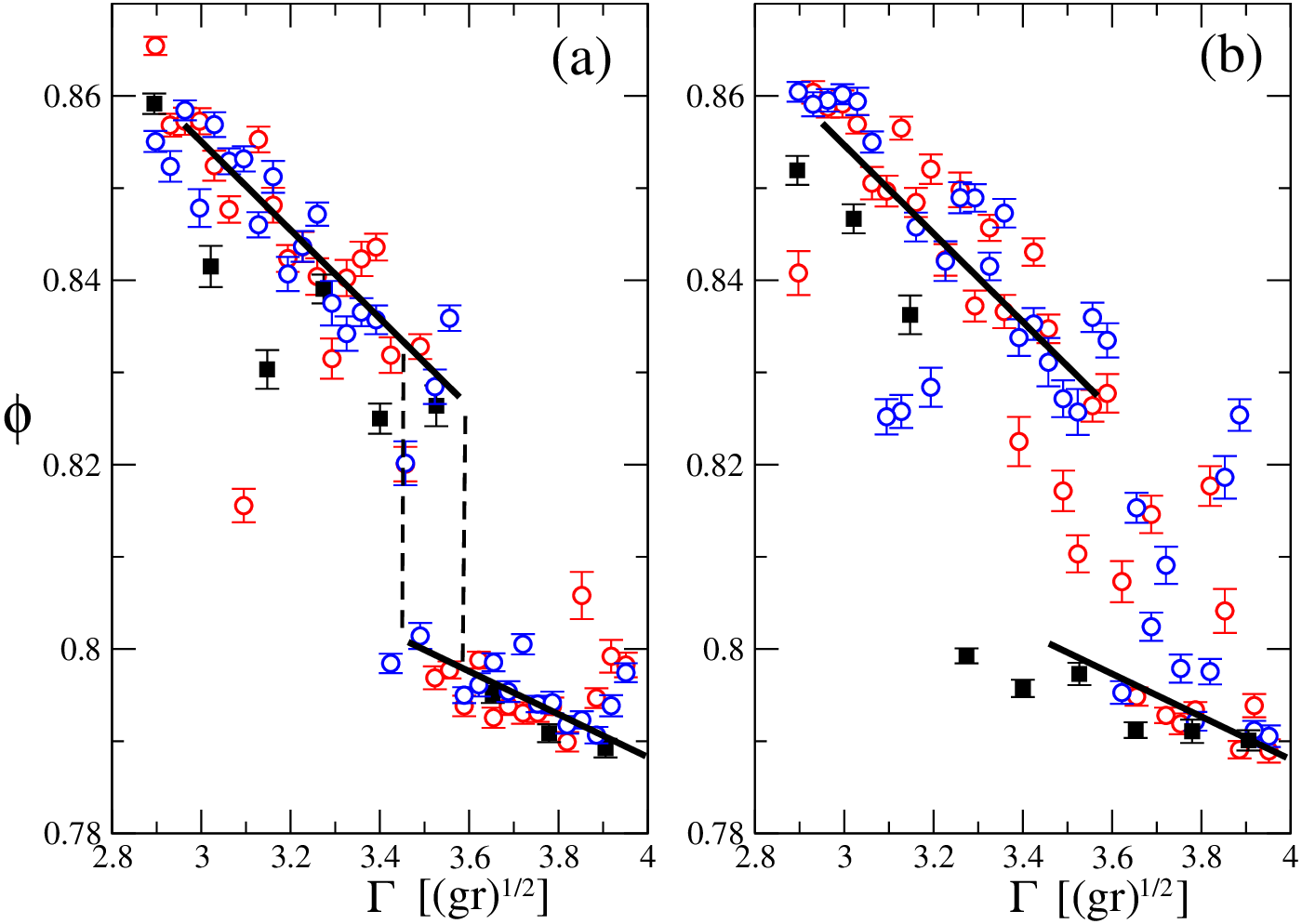}
 \caption{Mean packing fraction $\phi$ as a function of tapping intensity $\Gamma$ for tetrakaidecagons. Panel (a), increasing $\Gamma$. Panel (b), decreasing $\Gamma$. The red and blue data correspond to independent realizations of the tapping protocol. The black data correspond to the ones presented in Fig~\ref{fig3} for tetrakaidecagons, where larger steps in $\Gamma$ are taken. The full and dashed black lines are to guide the eye. Error bars as in Fig.~\ref{fig1}.}
 \label{fig4}
\end{figure}

We now focus on the behavior of polygons with larger number of vertices (from nonagons up to icosagons). Figure~\ref{fig3} shows the $\phi$--$\Gamma$ curves for each shape. One might have expected that a smooth change would appear in these curves as the number of vertices is increased up to a point where the behavior of the n-vertex polygon will converge to the one shown by disk packings. However, a sudden change is found as we move from dodecagons to triskaidecagons. While a continuous $\phi$--$\Gamma$ curve is observed for polygons with up to $12$ vertices, a sharp discontinuity in $\phi$ is present in all packings with polygons of $13$ vertices or more. A gap of ``forbidden'' values of $\phi$ appears between roughly $0.80$ and $0.83$ in all these polygon packings with more than $12$ vertices. It is important to mention that fluctuations are rather large, and configurations (microstates) with $0.80 < \phi < 0.83$ are rather common. It is the mean values that present a gap. 

A similar discontinuity has been seen in tapped disk packings simulated under a pseudo-dynamic algorithm \cite{pugnaloni3}. However, this is not observed in our simulations of disks (see brown data in the lower right panel in Fig.~\ref{fig3}) nor in previous molecular dynamic simulations where the same region of $\Gamma$ was explored \cite{arevalo}. The pseudo-dynamic algorithm \cite{pugnaloni3} conducts a deposition of disks that roll on top of each other without sliding. This might mimic, rather realistically, the behavior of regular polygons with a large number of vertices. These polygons behave like gears in the sense that they interlock very easily just as if they were infinitely rough disks. We presume this basic characteristic shared by polygons with many vertices and disk that roll without sliding is the underlying phenomenon that leads to the emergence of a discontinuous $\phi$--$\Gamma$ curve. We mention in pass that, although it is difficult to relate with the static packings studied here, a similar discontinuity has been reported in an oscillation experiment of a 2D granular sample \cite{shattuck}.

In order to have a rough indication of the nature of the transition, we have made a more detailed simulation for tetrakaidecagons ($N=14$). In Fig.~\ref{fig4}, the steady state value of $\phi$ is plotted for $\Gamma$ in the interval $[2.8,4.0]$ with a smaller $\Delta \Gamma$ step. In panel (a), we plot two independent experiments obtained by increasing $\Gamma$ alongside with the corresponding results from Fig.~\ref{fig3} (where a larger $\Delta \Gamma$ was used). The results for the reversed protocol in which $\Gamma$ is decreased is presented in panel (b) of Fig.~\ref{fig4}. In Fig.~\ref{fig4}(a), the system seems to present a first order type transition where metastable branches are explored. Since fluctuations are rather large for this small system sizes, the system may explore microstates compatible with both ``coexisting'' phases. Nevertheless, in Fig.~\ref{fig4}(b), where the protocol corresponds to decreasing $\Gamma$, the transition looks much smoother if the rate $\Delta \Gamma$ is reduced. Although the data is noisy, we can see that the width of transition region is rate dependent.

\subsection{Density fluctuations}
\label{fluctuations}

Density fluctuations have recently received renewed interest as a way to measure configurational temperature (as defined by Edwards \cite{edwards}) and entropy \cite{mcnamara}. It was in a fluidization experiment that a nonmonotonic dependence of the fluctuations $\Delta \phi$ as a function of $\phi$ in the steady state was first reported \cite{schroter}. In that work, Schroter \emph{et al.} found a minimum in the density fluctuations for spheres. However, a recent study on disks reported a maximum in fluctuations from both, experiments and simulations \cite{pugnaloni2}.  

\begin{figure}
 \centering
 \includegraphics[width=0.9\columnwidth]{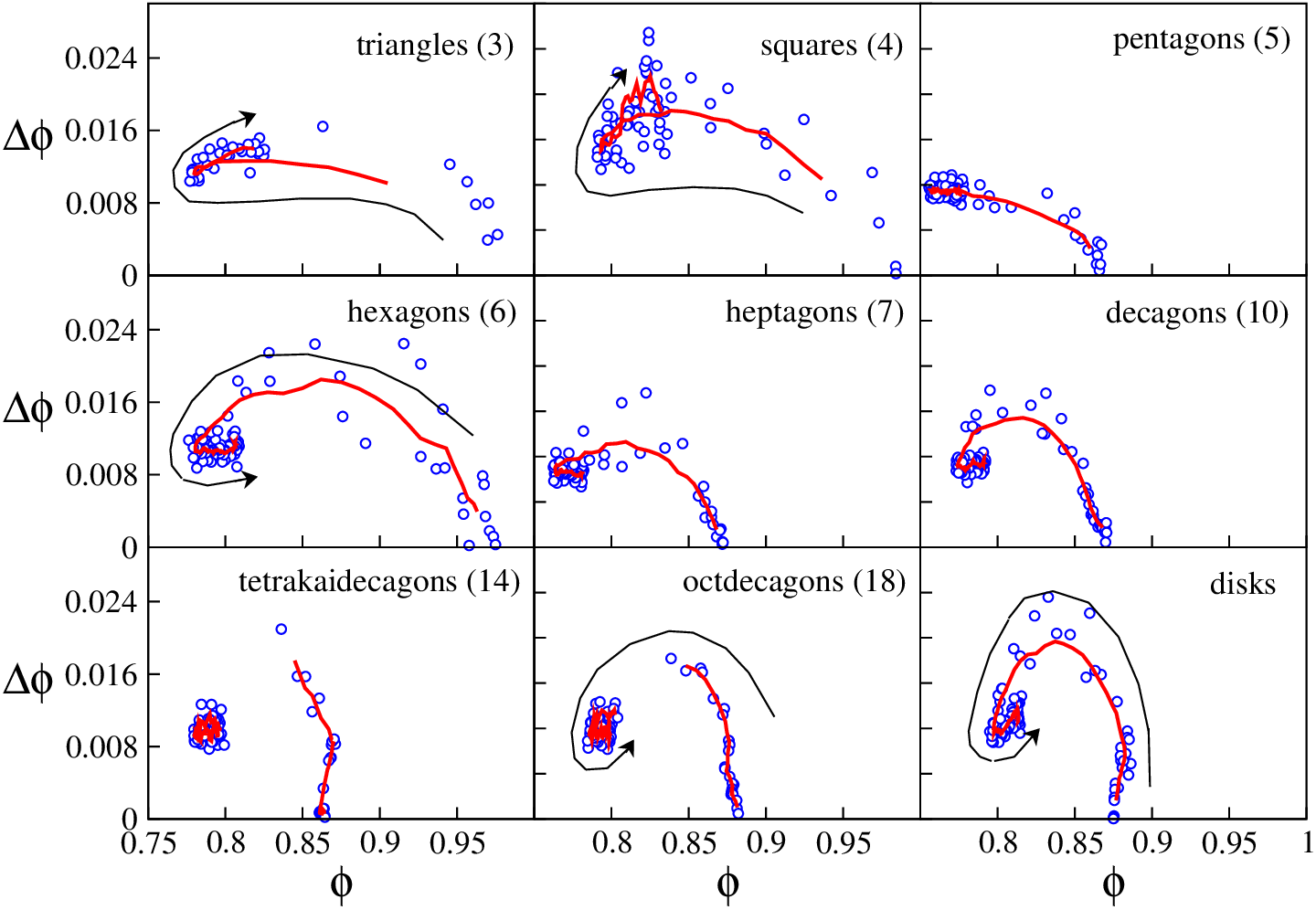}
 \caption{Standard deviation $\Delta \phi$ of the packing fraction in the steady state as a function of $\phi$. The red line is a simple running average to guide the eye. The arrows indicate the direction of increasing $\Gamma$.}
 \label{fig5}
\end{figure}

In Fig.~\ref{fig5} we show the steady state density fluctuations $\Delta \phi$ as measured by the standard deviation as a function of $\phi$ for several polygons and disks. The results for disks are entirely in agreement with Ref.~\cite{pugnaloni2}. A clear maximum in $\Delta \phi$ appears for disks. One can also see that states of equal $\phi$ at each side of $\phi_{min}$ present slightly different fluctuations. This indicates that these states are not equivalent and that $\phi$ is not sufficient to characterize the macroscopic state. A more detailed analysis of this can be found in Ref.~\cite{pugnaloni2} where the force moment tensor is found to be a suitable extra macroscopic variable in accordance with theoretical suggestions \cite{blumenfeld}.

The behavior of the density fluctuations in the polygon packings show the signal of the transition for shapes with $N>12$. However the same general trends as those seen for disks are observed. Interestingly, a peculiar behavior appears for triangles, squares and pentagons. Pentagons present the same fluctuations at both sides of the $\phi$ minimum whereas triangles and squares present a reversed situation where fluctuations are larger for large $\Gamma$, instead of smaller as seen in all other shapes. This change in trend should have an important impact in the calculation of configurational temperature and entropy. We will pursue this point further elsewhere.
 
\section{Conclusions}
\label{conclusions}
We have carried out simulations of the tapping of assemblies of regular polygonal grains and studied the steady state of such systems. The comparison with more widely studied disk packings has shown some general similarities but also remarkable new phenomenology.

On the one hand, beyond the expectable result for triangles, squares and hexagons that cover the space if gently tapped, polygons with $N>12$ show a sharp transition with a clear density gap. On the other hand, triangles and squares present density fluctuations that are larger at large tapping intensities in contrast with all other shapes (including disks).

A number of questions arise from this study that can lead future research. Some of these questions are: 

\begin{enumerate}
\item What is the true nature of the transition for polygons with a large number of vertices? Can this transition be effectively found in infinite rough disks? Can the low density coexisting phase be related with the so called \emph{random close packing} state \cite{radin,aristoff,jin,limon}.

\item Given that fluctuations have a different trend, is the granular (configurational) temperature in the case of triangles and squares radically different from that of other shape packings?

\item  Given that for pentagons the fluctuations are equivalent for states at each side of $\phi_{min}$ obtained with different $\Gamma$, which suggest that the states are equivalent, is the force moment tensor equivalent?
 
\end{enumerate}

\ack
We thank Ana Mar\'{\i}a Vidales and Irene Ippolito for valuable discussions. This work has been supported by CONICET and ANPCyT (Argentina).

\end{document}